# Asymmetric currents in a donor (D)-bridge (B)-acceptor (A) single molecule – revisit of the Aviram-Ratner diode

Haiying He and Ravindra Pandey[*]
Department of Physics and Multi--scale Technology Institute,
Michigan Technological University, Houghton, MI 49931

Govind Mallick and Shashi P. Karna[*]
US Army Research Laboratory, Weapons and Materials Research Directorate,
ATTN: AMSRD-ARL-WM, Aberdeen Proving Ground, MD 21005-5069

**Abstract**
The quantum transport via a donor (D)-bridge (B)-acceptor (A) single molecule is studied using density functional theory in conjunction with the Landauer-Büttiker formalism. Asymmetric electrical response for opposite biases is observed resulting in significant rectification in current. The intrinsic dipole moment induced by substituent side groups in the molecule leads to enhanced/reduced polarization of the system under a forward/reverse applied potential, thus asymmetry in the charge distribution and the electronic current under bias. Under a forward bias, the energy gap between the D and A frontier orbitals closes and the current increases rapidly; whereas under a reverse bias, the D-A gap widens and the current remains small.

[*]*Corresponding Authors.*
Ravindra Pandey: *pandey@mtu.edu*; Tel: +1 (906) 487-2086, Fax: +1 (906) 487-2933;
Shashi P. Karna: *skarna@arl.army.mil*; Tel: +1 (410) 306-0723, Fax: +1 (410) 306-0723.

## 1. Introduction

Following the seminal work by Aviram and Ratner[1] on the possibility of single-molecule diode, there have been a number of theoretical and experimental studies on the electron transport through organic molecules with a donor(D)-bridge(B)-acceptor(A) architecture in contact with semi-infinite metallic electrodes at the two ends.[2-10] The mechanism of the Aviram-Ratner (AR) diode consisting of D-B($\sigma$-type)-A can be summarized as a two-step process under a forward bias:

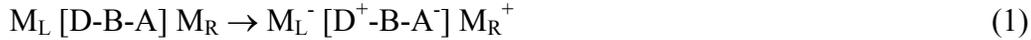
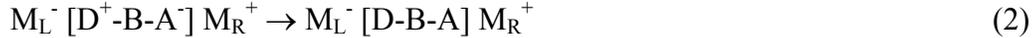

$$M_L [D-B-A] M_R \rightarrow M_L^- [D^+-B-A^-] M_R^+ \qquad (1)$$
$$M_L^- [D^+-B-A^-] M_R^+ \rightarrow M_L [D-B-A] M_R^+ \qquad (2)$$

In eq. (1), an electron tunnels from the right electrode $M_R$ to A, while an electron tunnels from D to the left electrode $M_L$, leading to the zwitterionic excited state $D^+-B-A^-$. It is followed by a reversion to the neutral ground state D-B-A via an intramolecular charge transfer (eq. (2)). The ultimate result is that one electron is moved from $M_R$ to $M_L$ in this process. Following a similar logic for the reverse bias process, however, it requires that D works as an acceptor and A as a donor, essentially requiring a high-energy $D^--B-A^+$ structure. A high energy barrier to the excited state $D^--B-A^+$ in the reverse bias process limits the electron flow from A to D. Thus the unidirectional current flow, i.e. the magnitude of current is much higher in the process involving $D^+-B-A^-$ (forward bias) state than in the $D^--B-A^+$ (reverse bias) state of the molecule. The $D^+-B-A^-$ excited state generally lies about 1-2 eV above the D-B-A ground state. The interplay of these two states is the foundation of the AR model for a unimolecular diode.

The tenet of the AR proposal is to illustrate the possibility of making molecular electronic units with the same functionalities as semiconductor devices. A donor (D) gives away some of its electrons playing the role of an n-type semiconductor, while an acceptor (A) withdraws weakly-bound electrons from other atoms in the molecular structure, playing the role of a p-type semiconductor. The energy barrier (B) unit, originally proposed in the AR model plays the role of a dielectric. Thus assuming that the B unit could be any spacer, including a $\pi$-electron conjugated unit, the D-B-A type of single molecule thus works as a p-n junction diode.

While single molecule architectures have been shown to exhibit current rectification in several experiments,[2-6] more recent theoretical studies have been unsuccessful in observing a discernible rectification effect in the AR-type D-B-A architecture.[7-9] The failure to observe the rectification in D-B-A architecture from theoretical calculations has led some authors to propose that the experimental observations of the unidirectional current in D-B-A architecture[7] results from the asymmetric coupling between the molecule and the electrodes. In the Langmuir-Blodgett experiment,[7] for example, an aliphatic chain is attached to one end of the molecule, while in the STM experiment, one of the electrodes – the STM tip is only weakly coupled to the molecule.[8] While the current experimental techniques have not reached the level of a single molecule measurement without some external perturbations which limit their applicability in understanding the underlying physics, it is possible to develop the fundamental physics from reliable first-principles calculations.

Here we present the results of our study on the electronic structure mechanism of molecular rectification in a D-A substituted para-terphenyl system. The present system differs from those of the AR model in the sense that the bridging moiety (B) is not a high-energy barrier $\sigma$-bonded structure. Rather we focused on a $\pi$-conjugated system to investigate the role of the charge distribution, presence of a permanent dipole moment due to the charge-separating D and A functional groups, and the effects of the external electric field on the energy-levels of the participating frontier orbitals. Another feature of our study is that the molecule is symmetrically and weakly coupled to the two electrodes without forming covalent bonds. This allowed us to focus on the electronic structure effects in molecule by conveniently isolating the molecular one-electron levels. We propose a new mechanism based on the fundamental property of a D-A substituted molecule – asymmetry in charge distribution and the permanent dipole moment. Our study shows that an electron-transporting dipolar molecule will always exhibit an asymmetric current under opposite biases while the magnitude of rectification in current depends on the details of the molecular electronic structure including spatial distribution of frontier orbitals, the effect of the external field on the frontier orbital charge distribution and eigenvalue variation of these orbitals under the external potential.

## 2. Calculations

The molecular architecture used in our study is shown in Figure 1. Specifically, the architecture consists of a para-terphenyl system, with 1,3-diaminobenzyl unit used as the donor (D) moiety and 1,2,4,5-tetracyanobenzyl as the acceptor (A) moiety bridged together by the middle phenyl ring. As a control, we also performed calculations on a para-terphenyl (TP) molecule, shown in Figure 1(b). Thus the D-A-B architecture and the control molecule both have the same π-conjugated backbone, differing only in the presence/absence of the donor (-NH$_2$) and acceptor (-CN) functional groups at the outer phenyl rings.

We used Au as electrodes, which seems to be a common choice for both experiments and theoretical studies. Despite the fact that the metal electrode-organic molecule interface plays an important role in molecular-scale electronics, in the current study, however, we focus on the roles played by the molecule itself. Further, in order to reduce the interface effects, we chose not to use the thiol group as the anchor of the molecule to the electrode, which is known[11,12] to induce a large interfacial and intermolecular charge redistribution and is involved in the formation of frontier orbitals of the molecular system. Instead, we use the C≡CH group to terminate the molecule at the end. The H-terminated geometry is analogous to the Langmuir-Blodgett (LB) film deposited on a metal substrate, where the molecule forms a week contact with the substrate surface. In our calculations, the H-Au distance was set to 2.0 Å, typical of the H-metal adsorption distance for several transition and post-transition metal surfaces.[13] The electrical response of a molecular device under bias was calculated in the presence of an external static electric field along the molecular chain. It was applied in both positive and negative directions to distinguish the forward (A→D or p-type→n-type) and the reverse (D→A or n-type→p-type) biases. The correspondence between the electric field and the source-drain bias potential is listed in Table I, in considering the distance between the left and right electrode to be around 23.3 Å. A two-atom Au chain, with Au-Au separation of 2.90 Å (the bond length in Au bulk is 2.885 Å), was used as the left and right metal contacts.

The electronic structures of the extended molecule Au$_2$-DBA-Au$_2$ (the center scattering region) under bias were calculated in the framework of the density functional theory (DFT) with B3LYP functional form[14,15] using Gaussian03 *ab initio* electronic structure code.[16] The LanL2DZ basis sets were used for all atoms. A symmetry-constrained geometry optimization was performed for the molecule at a zero field and used for all the calculations in presence of the electric field, since the geometry change in the molecule under an external electric field was found to be insignificant.[17]

The reliability of the quantum mechanical method itself is important in assessing the accuracy of the predicted results. Toward that, in the past decade, DFT has been considerably improved and with inclusion of appropriate XC functional, has been reliably used for studying, molecular electron transport.[18]

The current (*I*) - voltage (*V*) calculations were performed using the Green's function-based Landauer-Büttiker formalism.[19-21] The core scattering region was simulated by the extended molecular complex where atomic scale gold contacts were used for the molecule. And the semi-infinite effect of the left and right electrodes is included by the self-energy term $\Sigma$. Note that the coupling between Au contact and H (the termini of the molecule) is not strongly dependent on the size of the contact, thereby introduces negligible effect on the self energy. Additional details of the calculations can be found elsewhere.[22,23]

The total tunneling current of the system under a bias of *V* can be written as

$$I = \frac{2e}{h} \int_{-\infty}^{\infty} dE\, T\,(E,V)[f(E-\mu_1) - f(E-\mu_2)] \quad (3)$$

where $\mu_1$ and $\mu_2$ are the electrochemical potentials in the two electrodes under an external bias *V*, *f(E)* is the Fermi-Dirac distribution function. *T(E,V)* is the electron transmission function, which can be calculated from a knowledge of the molecular energy levels and their coupling to the metallic contacts as

$$T(E,V) = \text{Tr}\left[\Gamma_L(E)G(E,V)\Gamma_R(E)G^+(E,V)\right] \quad (4)$$

where the trace runs over all the orbitals and $\Gamma_{L(R)}(E)$ are twice the imaginary part of the self-energy matrices $\Sigma_{L(R)}(E)$. The Green's function which appears in eqn. (4) can be evaluated as

$$G(E,V) = \left[(ES - F) - \Sigma_L(E) - \Sigma_R(E)\right]^{-1} \quad (5)$$

where $F$ and $S$ are the Fock matrix and the overlap matrix (corresponding to a non-orthogonal set of wavefunctions) from the self-consistent Kohn-Sham solution of the extended molecule electronic-structure calculation for each bias, respectively.

### 3. Results and Discussion

The calculated current for this D-B-A molecular system is plotted in Figure 2. A significant rectification in the current is observed. As a figure of merit, the rectification ratio is defined as $\frac{I^+ - I^-}{I^+ + I^-}$, where $I^+$ and $I^-$ are the currents under the forward and the reverse bias, respectively. The calculated values are illustrated in the inset of Figure 2. The rectification factor increases continuously with an increase in the bias voltage, and reaches 0.94 at a bias of 1.2 V. In contrast, the corresponding *I-V* curve for the symmetric p-terphenyl molecule, as expected, shows highly symmetric characteristics.

In Table 1, we list changes from several aspects in the presence of an external bias potential. The dipole moment, a measure of charge separation in the extended molecule, varies considerably and almost linearly with the external field. The permanent dipole moment in this molecule due to the presence of the donor/acceptor side groups produces an internal electric field, with a net direction from A to D. The internal field strength is about the same order of magnitude as the external source-drain field. When the applied field is parallel to the internal field, the molecule experiences an enhanced polarization compared to when the external field direction is reversed. This leads to asymmetry in polarization and in the magnitude of the current under applied bias. The variation in the electrostatic potential profile due to the intramolecular charge redistribution is shown in Figure 3. In contrast, the TP molecule with no intrinsic dipole moment at a zero field, polarizes symmetrically under an external field applied in either direction.

The molecular orbitals due to the donor and the acceptor are the frontier orbitals near the Fermi level (as shown in Figure 4). Excluding the pure metallic orbitals from Au (LUMO and LUMO+1/LUMO+2), which do not contribute to the conductance of the molecule, the molecular orbitals from the D group are on the HOMO side and the molecular orbitals from the A group are on the LUMO side, due to the lower ionization potential of the D group. We note a continuous increase in the eigenvalue of the D orbital (HOMO) and a decrease in the A orbital (LUMO+2 for E < $3.09 \times 10^8$ V/m, or LUMO+1 for E > $3.09 \times 10^8$ V/m due to the band crossing) when the electric field varies from $-5.14 \times 10^8$ to $5.14 \times 10^8$ V/m. Therefore, the D-A orbital gap decreases with a positive applied field (along the A-D direction, i.e. p-n direction, forward bias) and increases under a negative field (reverse bias). The value of the $\Delta(\varepsilon_A - \varepsilon_D)$ at each value of the external field is listed in Table 1. Overall, our results qualitatively agree with those of ref. [9] in the low bias range considered. However, our results differ from ref. [10], which reported a narrowing of the HOMO-LUMO gap in both field directions. It is worth noting that despite a narrowing in either direction reported in ref. [10], the magnitude of the change due to the positive and negative biases are different. This difference turns out to be critical in observing asymmetric current for the two opposite biases.[10]

Nevertheless, this D-A gap does not close up to a positive bias of 5 V, suggesting a high barrier to the zwitterions excited state $D^+$-B-$A^-$, which implies that the Aviram-Ratner diode model cannot apply in this case. Furthermore, the pseudo Fermi level remains more or less the same, due to the Fermi pinning by the Au orbitals lying in the D-A gap.

We note that in the low bias regime, the dominant contribution to molecule conductance comes from the D orbitals (HOMO side), as the A orbitals are far above the pseudo Fermi level. Our analysis of the HOMO molecular orbital, which is attributed to the donor (D) group, shows that its wave character changes very little with the electric field in the range considered here. So does the character of the A orbital on the LUMO side. This observation again emphasizes the difference in electron transport mechanism in the present system and that in the AR proposal. Instead, the position of the HOMO gets closer to the pseudo Fermi level as the positive electric field (forward bias) increases, whereas it gets further away from the pseudo Fermi level with increasing the negative field (reverse bias). The calculated transmission functions are shown in Figure 5 for a series of biases. As a forward source-drain bias increases from 0.48, 0.96, to 1.20 V, the first transmission peak below the Fermi level due to the D orbital (located at -0.78, -0.61 and -0.52 eV) shifts towards the Fermi level, the same as the HOMO, whose separation from the Fermi level varies from -0.45, -0.30 to -0.22 eV (Table 1). The mismatch of the exact values between the transmission peak position and the molecular orbital energy level of the extended molecule $Au_2$-DBA-$Au_2$ on the energy scale, however, is due to the fact that two additional terms – self energies $\Sigma_{L(R)}(E)$ describing the semi-infinite effect of the left and right electrodes, were added to the Hamiltonian of this open transport system Au(left)-DBA-Au(right) in the calculation of electron transport. Similarly, the first transmission peak above the Fermi level due to the A orbital shifts in the same fashion as the LUMO A orbital. It is clearly seen that the resonance peak in transmission function due to the HOMO (or D) orbital is far away from the bias window at a bias of 0.48 V; approaches closer to the bias window at a bias of 0.96 V, and the tail of that peak starts to contribute to current; and falls into the bias window at a bias of 1.20 V, giving rise to a dramatic increase in current. The current is rather flat under the reverse bias within 1.00 V, and starts to lift up a bit at a bias of 1.20 V. This explains the current rectification in the present architecture. In contrast, the D-A gap for the TP molecule, which lacks intrinsic dipole moment, changes symmetrically under a forward/reverse bias.

It is worth noting that the S atoms used for Au-molecule attachment in other studies[9, 10] reduces the dipole moment and smears out the asymmetry of charge distribution, thereby reducing the magnitude of rectification of the overall molecular system. The reason is that the S atom also acts as a good electron donor.[24] Thus the presence of the S atom as a molecule-metal binder has a strong influence on not only on the molecule-electrode chemical bonding but also on the intramolecular charge redistribution.

Finally, we would like to point out that the current study is based on an assumption of the symmetric potential drop on both electrode sides. The rectification effect we calculate results from the polarity and therefore the dipole moment of the molecule, which along with the one-electron frontier orbital energy levels, experience enhanced modulation in the presence of an external electric field, leading to a unidirectional current flow. The energy level mechanism of the current rectification is shown in Figure 6. In the energy diagrams, we have drawn the frontier orbitals due to the D and A groups under zero and non-zero electric fields corresponding to the forward and reverse biases, as well as the bias window. At zero bias, the HOMO, due to the D group is much closer to the pseudo Fermi level. Under a forward bias, the D-A gap shrinks and the D orbital moves into the bias window contributing to the total current; whereas under a reverse bias, the D-A gap widens and both the D and A orbitals move further away from the bias window. The different response under opposite applied fields, leads to the rectification effect through the molecule.

### 4. Summary

In summary, the electron transport properties of a DBA model system have been investigated in the frame work of Landauer-Buttiker formalism using DFT and Green's function methods. The diode behavior is observed in the absence of redox reaction (zwitterionic state). The electrical rectification is due to its intrinsic dipole moment, which makes the molecule respond differently when the direction of the external electric filed changes. Under a forward applied potential (in the same direction as that of the intrinsic dipole moment), the energy gap between the D and A frontier orbitals closes and the electronic current increases rapidly; whereas under a reverse applied potential, the D-A energy gap widens and the current remains small. This leads to the asymmetry in the *I-V* curve.

It is also important to note that the effectiveness of the rectification process depends on the distribution of the frontier molecular orbitals arising from the D and A side groups as well as on the polarizability of the system under an electric field. If the HOMO-LUMO gap of the system is too large, a significant rectification in current may not be observed until a high bias voltage is applied.[9] Similarly, if the system is good in screening the applied electric field and induces a relatively small change in the dipole moment, the rectification effect will again be small.[10, 25]

*Acknowledgement*-The work at Michigan Technological University was performed under support by the DARPA through contract number ARL-DAAD17-03-C-0115. The work at Army Research Laboratory (ARL) was supported by the DARPA MoleApps program and ARL-Director's Research Initiative-FY05-WMR01. Helpful discussions with S. Gowtham, K. C. Lau and R. Pati are acknowledged.

**References**

1. Aviram A.; Ratner, M. A. *Chem. Phys. Lett.* **1974**, *29*, 277.
2. Geddes N. J.; Sambles J. R.; Jarvis D. L.; Parker W. G.; Sandman D. J. *Appl. Phys. Lett.* **1990**, *56*, 1916.
3. Pomerantz M.; Aviram A.; McCorkle R. A.; Li L.; Schrott A. G. *Science* **1992**, *255*, 1115.
4. Martin A. S.; Sambles J. R.; Aswell G. J. *Phys. Rev. Lett.* **1993**, *70*, 218.
5. Metzger R. M.; Chen B.; Höpfner U.; Lakshmikantham M. V.; Vuillaume D.; Kawai T.; Wu X.; Tachibana H.; Hughes T. V.; Sakurai H.; Baldwin J. W.; Hosch C.; Cava M. P.; Brehmer L.; Ashwell G. J. *J. Am. Chem. Soc.* **1997**, *119*, 10455.
6. Metzger R. M.; Xu T.; Peterson I. R. *Angew. Chem. Int. Ed. Engl.* **2001**, *40*, 1749.
7. Krzeminski C.; Delerue C.; Allan G.; Vuillaume D. *Phys. Rev. B* **2001**, *64*, 085405.
8. Taylor J.; Brandbyge M.; Stokbro K. *Phys. Rev. Lett.* **2002**, *89*, 138301.
9. Stokbro K.; Taylor J.; Brandbyge M. *J. Am. Chem. Soc.* **2003**, *125*, 3674.
10. Staykov A.; Nozaki D.; Yoshizawa K. *J. Phys. Chem. C* **2007**, *111*, 11699.
11. Yaliraki S. N.; Kemp M.; Ratner M. A. *J. Am. Chem. Soc.* **1999**, *121*, 3428.
12. Lang N. D.; Kagan C. R. *Nano Lett.* **2006**, *6*, 2955.
13. Lamoen D.; Ballone P.; Parrinello M. *Phys. Rev. B* **1996**, *54*, 5097.
14. Becke A. D. *J. Chem. Phys.* **1993**, *98*, 5648.
15. Lee C.; Yang W.; Parr R. G. *Phys. Rev. B* **1998**, *37*, 785.
16. Frisch M. J. *et al.*, GAUSSIAN03, Gaussian, Inc., Pittsburgh PA, 2003.
17. For instance, under an electric field with a strength of $5.14 \times 10^8$ V/m, the maximum variance in the bond length is ~0.005 Å, in the bond angle ~1° and in the dihedral angles between two adjacent benzene rings ~2°.
18. Zhang G.; Musgrave C. B. *J. Phys. Chem. A* **2007**, *111*, 1554.
19. Landauer R. *J. Phys.: Condens. Matter* **1989**, *1*, 8099.
20. Büttiker M. *Phys. Rev. Lett.* **1986**, *57*, 1761.
21. Datta S. *Electronic transport properties in mesoscopic systems*, Cambridge University Press, Cambridge, 1995.
22. Tian W.; Datta S.; Hong S.; Reifenberger R.; Henderson J. I.; Kubiak C. P. *J. Chem. Phys.* **1998**, *109*, 2874.
23. He H.; Pandey R.; Karna S. P. *Chem. Phys. Lett.* **2007**, *439*, 110.
24. Dhirani A.; Lin P. H.; Guyot-Sionnest P.; Zehner R. W.; Sita L. R. J. *Chem. Phys.* **1997**, *106*, 5249.
25. Elbing M.; Ochs R.; Koentopp M.; Fischer M.; von Hänisch C.; Weigend F.; Evers F.; Weber H. B.; Mayor M. *Proc. Nat. Acad. Sci. USA* **2005**, *102*, 8815.

TABLE 1 Summary of the dipole moments, the energy gap between the D/A frontier orbitals $\Delta(\varepsilon_A-\varepsilon_D)$ and the HOMO (due to the D group) energy with respect to the pseudo Fermi level $\Delta(\varepsilon_D - E_F)$ of the extended molecule under a series of applied electric fields. The corresponding applied potentials are also calculated from the knowledge of the distance between two electrodes.

| field ($10^8$ V/m) | applied potential (V) | dipole (Debye) | $\Delta(\varepsilon_A-\varepsilon_D)$ (eV) | $\Delta(\varepsilon_D - E_F)$ (eV) |
|---|---|---|---|---|
| -5.14 | -1.20 | -0.15 | 2.20 | -0.50 |
| -4.11 | -0.96 | 0.45 | 2.12 | -0.54 |
| -3.09 | -0.72 | 1.05 | 2.04 | -0.58 |
| -2.06 | -0.48 | 1.66 | 1.96 | -0.62 |
| -1.03 | -0.24 | 2.26 | 1.88 | -0.66 |
| 0.00 | 0.00 | 2.87 | 1.80 | -0.61 |
| 1.03 | 0.24 | 3.48 | 1.66 | -0.53 |
| 2.06 | 0.48 | 4.09 | 1.64 | -0.45 |
| 3.09 | 0.72 | 4.70 | 1.56 | -0.38 |
| 4.11 | 0.96 | 5.32 | 1.48 | -0.30 |
| 5.14 | 1.20 | 5.94 | 1.40 | -0.22 |

**Figure Captions:**

Figure 1 Schematic illustrations of the extended molecules (a) $Au_2$-DBA-$Au_2$, (b) $Au_2$-TP-$Au_2$ coupled to the left and right electrodes. The semi-infinite effect of the electrodes is included by the self-energy term $\Sigma$.

Figure 2 Current-voltage curves for DBA (in red solid circle and solid line) and TP (in green empty circle and dashed line). The calculated rectification factor for DBA is plotted in the inset.

Figure 3 Electrostatic potential for DBA under a bias of (a) -1.20 V, (b) 0.00 V and (c) 1.20 V (excluding the linear potential drop due to the external electric field). The green surfaces correspond to the regions with a positive electrostatic potential energy of 0.1 eV, whereas the purple area to the regions with a negative electrostatic potential energy of -0.1 eV.

Figure 4 Molecular spectra of the DBA extended molecule under a series of external electric field. The HOMO orbital due to the donor group (D) is in green and the first DBA virtual orbital due to the acceptor group (A) is in magenta. The pseudo Fermi level $E_F$ in each case is taken as the average between the HOMO (a D orbital, in green) and the LUMO (a Au orbital, in black) of the extended molecular system.

Figure 5 Transmission functions for DBA under a series of forward biases. The bias windows are shown in magenta lines. The pseudo Fermi level is aligned to zero.

Figure 6 Illustration of current rectification mechanism in a D-B-A type of molecule with weak donors and acceptors. The bias window is shown in light blue.

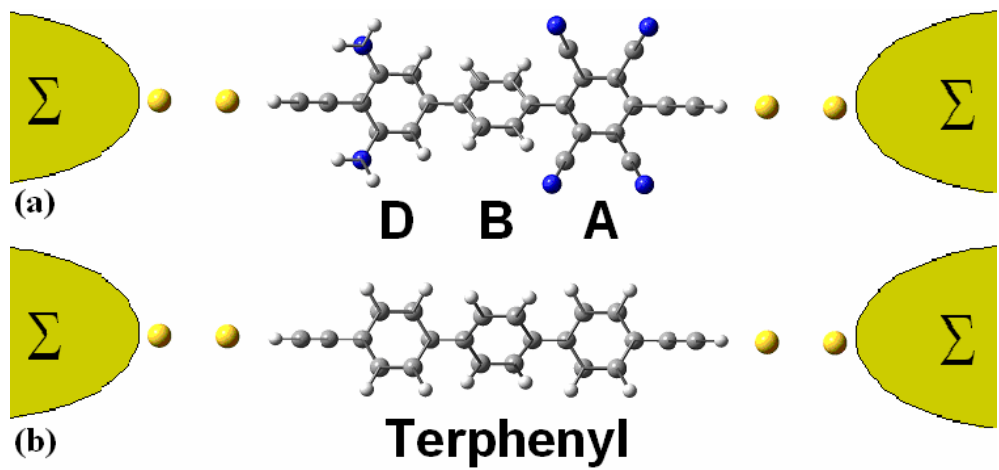

Figure 1 Haiying He *et al*.

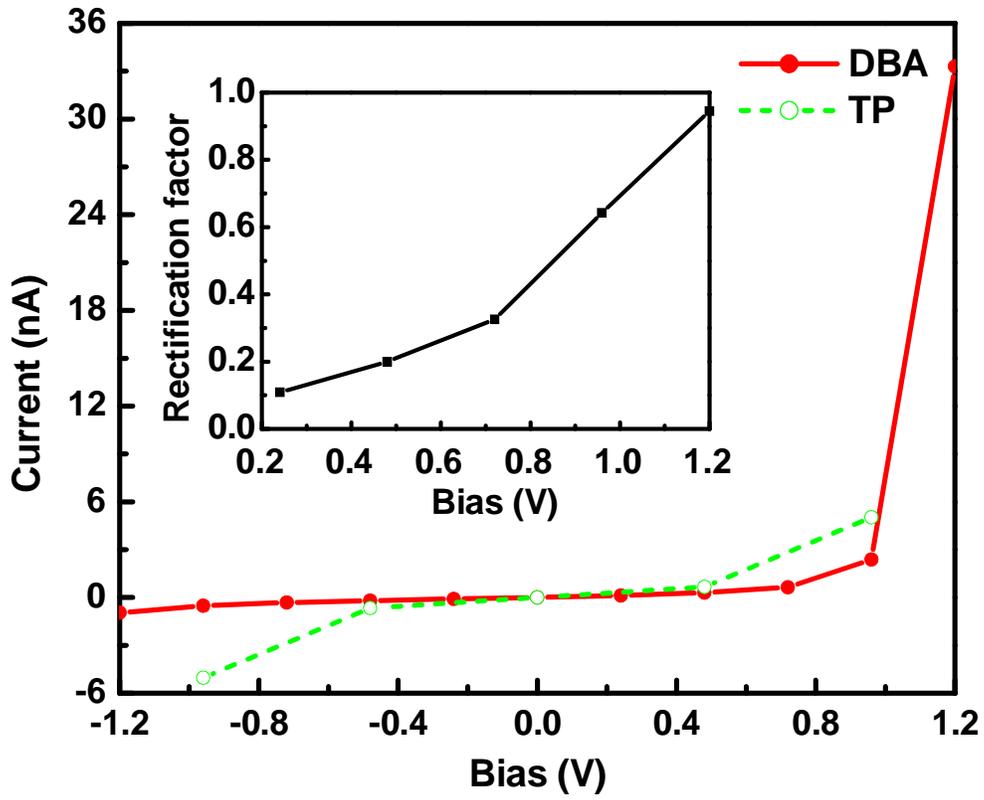

Figure 2 Haiying He *et al*.

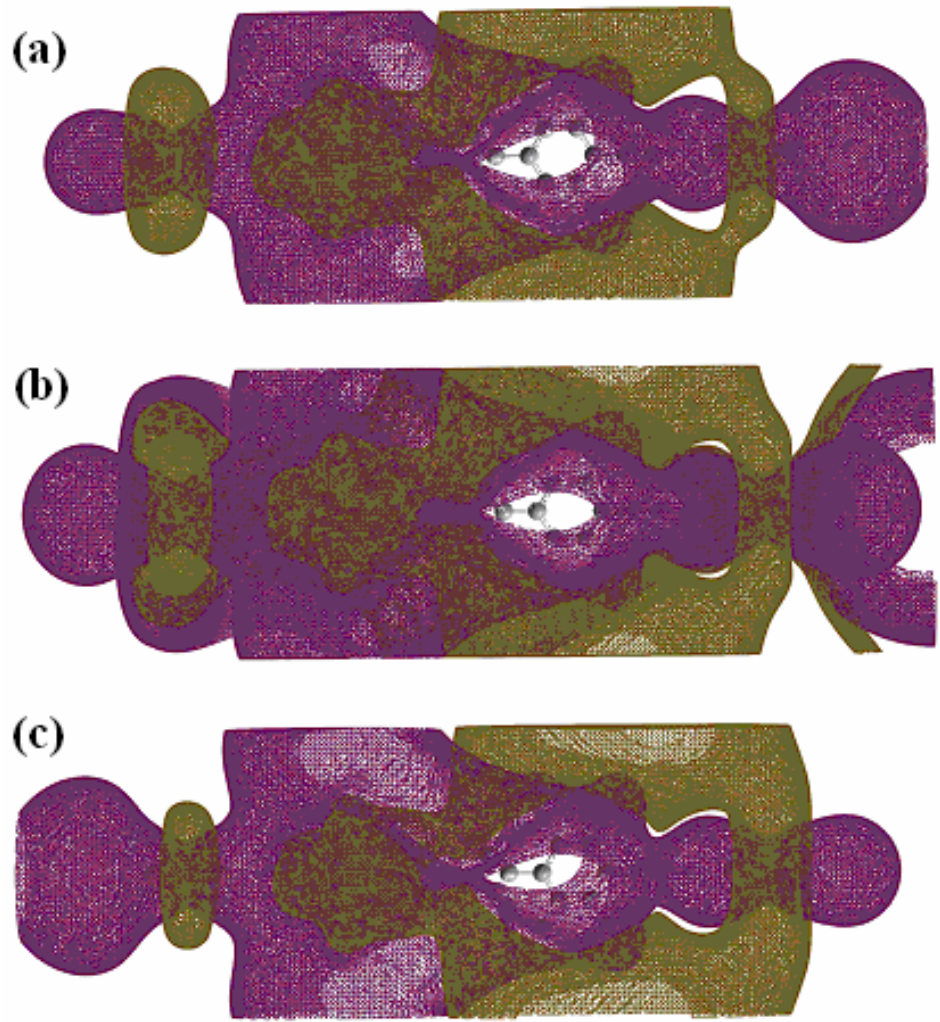

Figure 3 Haiying He *et al*.

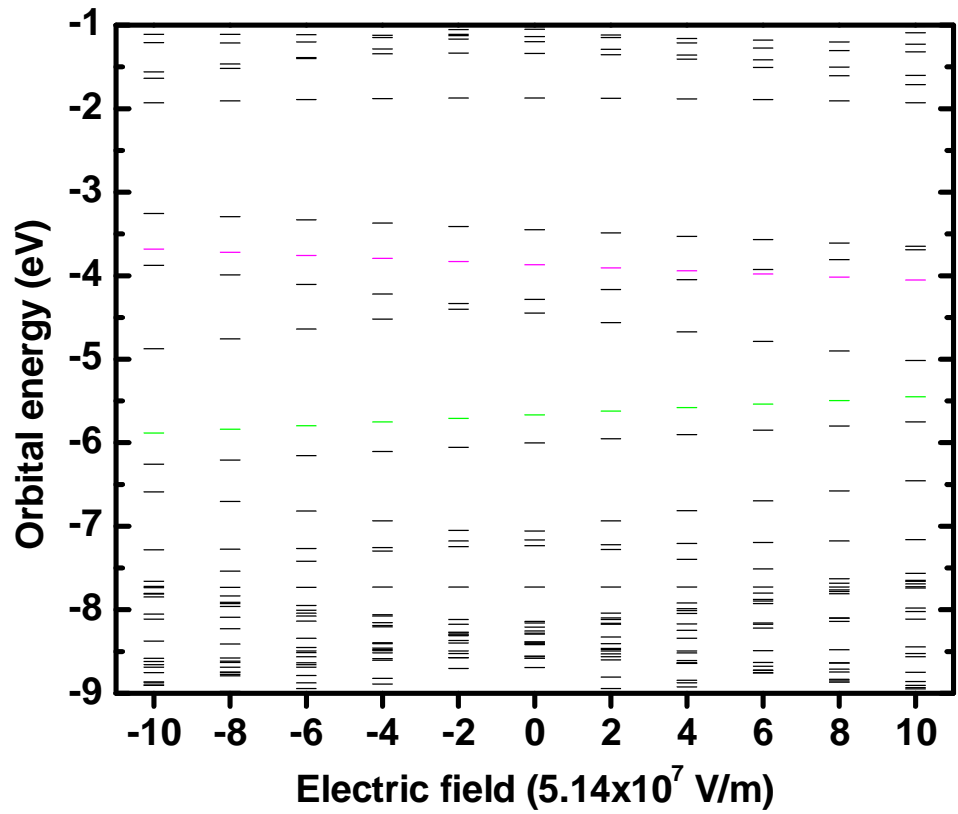

Figure 4 Haiying He *et al*.

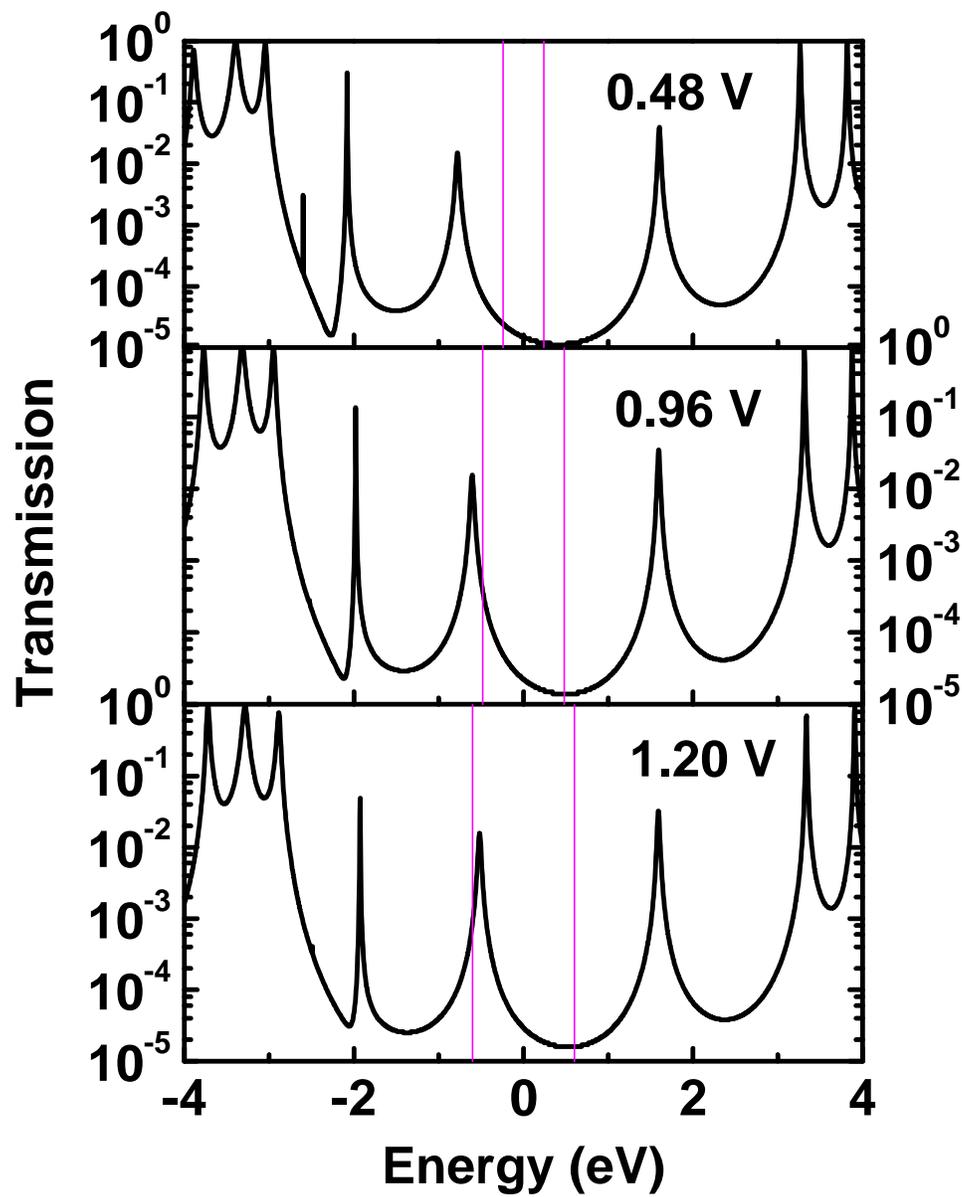

Figure 5 Haiying He *et al*.

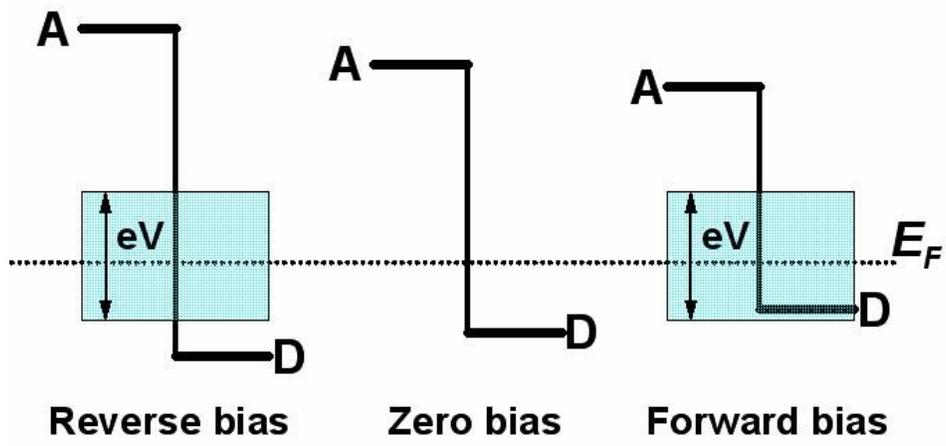

Figure 6 Haiying He *et al*.